\documentclass[10pt]{revtex4-2}

\usepackage{amsmath,}
\usepackage{graphicx}

\begin{document}

\title{On Three ``Anomalous'' Measurements of Nonlinear QPC Conductance}

\author{Mukunda P. Das$^1$ and Frederick Green$^2$}

\affiliation{$^1$ Department of Fundamental and Theoretical Physics,
RSP, The Australian National University, Canberra, ACT 2601, Australia.}

\affiliation{$^2$ School of Physics, The University of New South Wales,
Sydney, NSW 2052, Australia.}

\begin{abstract}
Practical mesoscopic devices based on quantum point contacts (QPCs) must
function at operating point involving large internal driving fields.
Experimental evidence has accumulated to display anomalous nonlinear
features of QPC response beyond the capacities of accepted tunnelling-based
models of nonlinear quantum transport. Here we recall the physical setting
of three anomalous QPC experiments and review how, for two of them, a
microscopically based nonequilibrium quantum kinetic description – the
correct physical boundary conditions being crucial - has already overcome
the predictive limitations of standard nonequilibrium mesoscopic models.
The third experiment remains a significant challenge to all theorists.
\end{abstract}

\maketitle

\section{Introduction}

Phenomena in quantum transport unique to two-dimensionally constrained
electronic systems, formerly considered exotic, have since become
bread-and-butter items for condensed-matter physics. Already for some time,
in the form of quantum-well-confined structures, they have been deeply
embedded in the design even of prosaic consumer electronics.

Slightly later than the basic two-dimensional breakthroughs of the
1970s came those at the next level, for structures virtually,
if not literally, at one dimension (1D). Their fundamental exploration
is very much a current research focus for the physics of electronic
transport. While one might argue that their truly large-scale applications
have some way to go to outdo the widespread success of their two-dimensional
predecessors, it is taken for granted that this is a basic matter of
concerted technical development.

It is the physical underpinning of the 1D device realm that forms the
topic in the following, somewhat condensed, review. Readers who may want
more detailed information on all the accepted methodologies based on the
reservoir analogy, as well as background on our own application of
established quantum kinetics, are invited to consult the extended references
in the long survey paper
\cite{gen}
as well as in the more specialized 
\cite{2000},
\cite{ijmpb}
and
\cite{mesotrans}.

We concentrate on aspects of observed 1D transport behaviour that appear to
go against accepted theoretical models. Both experimentalists and theorists
tend to refer to these laboratory results as ``anomalous''; but any
consistent departure from experimental expectations will always be
perceived as abnormal -- until a tenable explanation emerges. The question is:
``anomalous'' relative to what? The answer is: relative to the received
theoretical understanding. Nature entertains no anomalies. It is we as
investigators who are subject to a momentary cognitive dissonance.
Anomalous results, by their very presence, tell us that
more is demanded of a quantitative theory than anything in common currency.

Anomalies in nonlinear 1D transport emerged around the turn of this
century. The actual surprise, there, is not their discovery but that
they seemed to remain in the ``too-hard basket'', practically unattended
to in the literature. Here we review two cases that have now been
addressed, and cite a third intriguing instance still defying analysis.

Our first example is excess 1D conductance in a working region where the
conductance quantum $e^2/\pi\hbar$ is supposed never to be exceeded
\cite{1,dp2};
the second is anomalous hot-electron noise in a 1D wire,
or quantum point contact (QPC)
\cite{2}.
The third striking example, which we can discuss only in passing, is
the nonlinear Aharanov-Bohm effect
\cite{3,3.1}.
These measurements, largely bypassed by theorists, supply appreciable
evidence for reconsidering the limits of accepted mesoscopic transport
models
\cite{4}.

\section{Background}

We begin by recapitulating the distinctive transport
behaviour of quasi-1D ballistic channels. Some thirty-five years ago
two measurements
\cite{vw,pep}
first demonstrated the quantization of 1D conductance
as a function of device carrier density,
in sequential steps close to
the universal quantum $G_0 = e^2/\pi\hbar$, termed the
Landauer conductance in this field (the same as
the quantum-Hall unit, up to a factor of two).
A weighty literature explaining the quantization
\cite{l1,l2,l3,l4}
and its extension to noise
\cite{bb}
was soon established to offer a readily
appealing, seemingly adaptable description.

\subsection{Standard account}

Consider two macroscopic metallic electron reservoirs, left and right,
at chemical potentials
$\mu_L \geq \mu_R$ and linked by a long, uniform narrow wire supporting
a series of discrete quantum levels, or channels,
produced by the lateral confinement defining the quasi-1D conductor.
The channels are assumed independent, each with its own set of
conductive 1D states able to transmit charge carriers. 

In the Landauer current formula
\cite{l3,l4}
the reservoirs have
equilibrium electron distributions
$f(\epsilon_k - \mu_L)$ and $f(\epsilon_k - \mu_R)$
where wave vector $k$ indexes the conductive states in the 1D band,
with $\epsilon_k$ and $v_k = \hbar^{-1}d\epsilon_k/dk$
the band energy and group velocity respectively.
Phenomenologically one may argue that the net current should be
a difference of right- and left-moving electron fluxes:
\begin{eqnarray}
I
=
J_R - J_L
= \int^{\infty}_0 \frac{dk}{\pi} (-e)v_k
{\Bigl( f(\epsilon_{k'} - \mu_R){\cal T}(\epsilon_{k'})
(1 - f(\epsilon_k - \mu_L))
\Bigr.}
\cr
~~~ ~~~ ~~~ ~~~ ~~~ ~~~ ~~~ ~~~ ~~~ ~~~ ~~~ ~~~ ~~~ 
{\Bigl.
- f(\epsilon_k - \mu_L){\cal T}(\epsilon_k)
(1 - f(\epsilon_{k'} - \mu_R)) \Bigr)}
\label{01}
\end{eqnarray}
where the right-side momentum $k'$ is defined via
$\epsilon_{k'} = {\rm max} \{0, \epsilon_k + \mu_R - \mu_L \}$
and ${\cal T}(\epsilon)$ a unitary transmission factor when
the wire presents a barrier to coherent transmission.
The equation assumes total statistical decoupling
of initial and final state occupancies. As such it is a
revisiting of older tunnelling prescriptions
\cite{frensley}
such as the early one of Bardeen and of Esaki's extension to the tunnelling
diode.

We remark that tunnelling formulae are appropriate between
autonomous reservoirs communicating solely by overlap of the
exponential tails inside the barrier region. The fundamentally
stochastic coupling of the left- and right-localized carrier
distributions, in two disjoint bands,
does not apply to metallic transport. In that case, by contrast,
the physics is dominated by the flux-bearing eigenstates.
These states are delocalized across the
entire participating system and support a single distribution (not two)
in a unified conduction band, with a quantum barrier or with none.

At equilibrium, with equal chemical potentials
$\mu_L = \mu = \mu_R$,
the current given by Equation (\ref{01}) is zero.
A small driving voltage $V$ applied right to left will
raise $\mu_L$ above $\mu_R$ by $eV$. Noting that
$v_kdk = d\epsilon_k/\hbar$, in the low-temperature limit
the current becomes
\begin{eqnarray}
I
&=&
\frac{e}{\pi\hbar}\int^{\infty}_0 d\epsilon {\cal T}(\epsilon)
{\Bigl( f(\epsilon - \mu_R - qV) -  f(\epsilon - \mu_R) \Bigr)}
\cr
\cr
&=&
-\frac{e^2V}{\pi\hbar}\int^{\infty}_0 d\epsilon {\cal T}(\epsilon)
\frac{\partial f}{\partial \epsilon} + {\cal O}(e^2V^2)
~\to~
G_0 {\cal T}(\mu)V
\label{02}
\end{eqnarray}
in which the conductance quantum now appears. One can repeat the
argument for a number of occupied discrete channels assuming
that that they are independent of one another, excluding any
cross-talk in the active region as well as any interaction
mediated by their common lead reservoirs. Under those terms the
total conductance becomes
\begin{eqnarray}
G = \frac{I}{V} = G_0 \sum_{E_n < \mu} {\cal T}(\mu - E_n)
\label{03}
\end{eqnarray}
with $E_n$ the discrete energy thresholds for the quantum-confined
1D conduction bands, or channels. This accounts for the
characteristic constancy of conductance as a function of
channel density when this is modulated by altering chemical
potential $\mu$, usually by gate control of the device.
Since ${\cal T}(\epsilon) \le 1$, it follows from Equation (\ref{03})
that no individual step can exceed $G_0$, the ideal maximum.

Efforts to generalize the basic formula, Equation (\ref{01}), to the
nonlinear regime have been made, and keep being made. Few, if any, of the
best advertised attempts see fit even to re-evaluate, let alone leave behind,
the idea of tunnelling within an environment that is clearly metallically
conductive. We have discussed elsewhere
\cite{gen},
in more formal detail, the most popular of these recent approaches:
the nonequilibrium Green function (NEGF) formalism.

In any of a variety of guises
\cite{paulsson,ahp}
NEGF computes the self-energy for the putative interacting Green function
(the purely single-particle version, that is) within some correlated electron
model in localized form. The result is a locally determined effective
transmission amplitude ${\cal T}(\epsilon)$ richer than the Landauer version
but which nevertheless -- based as it is on a one-body object -- is a
quasiparticle spectral weight that never exceeds the unitarity bound in
the way already observed experimentally
\cite{1,dp2}
and described in the following. NEGF still does nothing about the imaginative
but unfortunately inappropriate boundary conditions insisting on localized
tunnelling inside a fully delocalized, if non-uniform, integrally
metallic medium.

\subsection{Difficulties and remedies}

We need not dwell any more on the use of a tunnelling scenario in cases
where the physics of conduction is undoubtedly metallic, for there
are two other difficulties with Equation (\ref{03}) describing
conductance quantization. The more obvious issue is that conductance
implies ohmic resistance, which implies Joule heating. It is immaterial
whether the resistive device is macro-, meso- or nano-scopic. In distilling
Equation (\ref{01}) to get to (\ref{03}) there is no appeal
to the process of dissipation; coherent quantum transmission is
dissipationless as well as reversible (unlike Joule heating).

The second, less obvious problem, is the assumption of independent
conducting channels. All their  carriers are injected from, then
received by, bulk metallic leads. It is inevitable that there
must be an interplay between band populations in their transition
to and from the common leads.
This is likely to affect the details of the carriers' distribution
as some level,
even if they did not interact in their discrete channels while
crossing the active region. In the next Section we return to this
aspect in the context of measurements by de Picciotto {\em et al}
\cite{1,dp2}.
First we deal with the problem of dissipation.

Loss of electrical energy absorbed from driving fields is a nonequilibrium
effect that impacts directly upon the detailed physics of transport.
Dissipation always has an immediate role in any quantitative account
of what is, in the end, a case of ohmic conduction. As the
experiments surveyed below show, the problem becomes pressing well
away from linear response where, as is now acknowledged, coherent
transmission theory has less to offer in a systematic way.

A solution already exists within quantum kinetics taken in its
its long-wavelength limit, the quantum-Boltzmann equation,
and in the succinct form of the canonical Kubo-Greenwood relation
to which the kinetic equation conforms
\cite{mahan}.
We will not reproduce the full derivation of the
quantum kinetic formula that corresponds to Equation (\ref{03})
\cite{gen},
but give some sense of a different philosophy that leads to
an equivalent result while fully including the physics of dissipation.
We start with a single {\em metallic} channel.
If one takes the familiar Drude-Sommerfeld conductance formula in
the low-temperature limit, expressing its customary relaxation rates
in terms of mean free paths, one has
\begin{eqnarray}
G =
\frac{e^2 n}{m^*u_{\rm F}L}
\frac{\lambda_{\rm el}\lambda_{\rm in}}{\lambda_{\rm el}+\lambda_{\rm in}}.
\label{04}
\end{eqnarray}
Parameters are: the operative channel length $L$;
effective mass $m^*$ and Fermi velocity $u_{\rm F}$
for a quadratically dispersive conduction band; 1D carrier density
$n = 2k_{\rm F}/\pi$ in terms of $k_{\rm F}$, the Fermi wave vector;
$\lambda_{\rm el}$ and $\lambda_{\rm in}$ are the elastic and inelastic
scattering mean free paths respectively, the latter responsible
for resistive loss and, crucially, for stability of the steady state
\cite{gen}.

So far, the quantitative content is perfectly conventional.
The physical reasoning specific to a quasi-ballistic structure now follows.

Transport within the channel is ballistic and not necessarily coherent.
There do exist finite mean free paths but these are
not set by the usual considerations in the bulk, rather by the
carriers' interactions with the leads through stochastic injection
and extraction.
By the same token the operational length of the channel is also
set by the dynamics of its interfaces with the leads; for, at
mesoscopic scales, the geographic boundary between channel and leads
becomes ill-defined. Conceptually the limiting device boundary $L$
itself becomes identifiable with the maximum mean free path, the
scale beyond which any notion of ballisticity is lost.
(In transmission parlance this scale is the ``coherence length''.)

On these assumptions, in Equation (\ref{04}) we make the replacement
$L \to {\rm max}\{\lambda_{\rm in}, \lambda_{\rm el}\}$.
Rationalizing the factor $n/m^*u_{\rm F} = 2/\pi\hbar$, we arrive at
\begin{eqnarray}
G =
\frac{e^2}{\pi\hbar} {\left(
\frac{2{\rm min}\{\lambda_{\rm in}, \lambda_{\rm el}\}}
{\lambda_{\rm in}+\lambda_{\rm el}} \right)} \leq G_0.
\label{05}
\end{eqnarray}
There is no functional difference between this expression and
the lossless Equation (\ref{03}). Both have the same ideal maximum
$G_0$; Equation (\ref{05}) attains it when the mean free paths
are matched. The latter formula gives dissipation its proper role and
avoids the conceptual confusion of Equation (\ref{03}), which pits
the complete coherence of a fully delocalized transmitted flux against
the complete stochasticity of the localized static boundary conditions,
with no rational handle on energy loss.
The kinetic basis for Equation (\ref{05}) also
allows a systematic approach to current noise
\cite{gtd};
more on this below.

We come to the task of resolving the second and more
challenging issue presented by the hypothesis of
total independence for discrete channel populations.
When discrete channels are able to interact,
the possibility opens up for interband carrier transfer.
This immediately turns the transport calculation into a
correlated many-body one; a case in which unitarity in each channel fails
and a collective description is required.

\section{Interacting one-dimensional bands}

\centerline{
\includegraphics[width=6truecm]{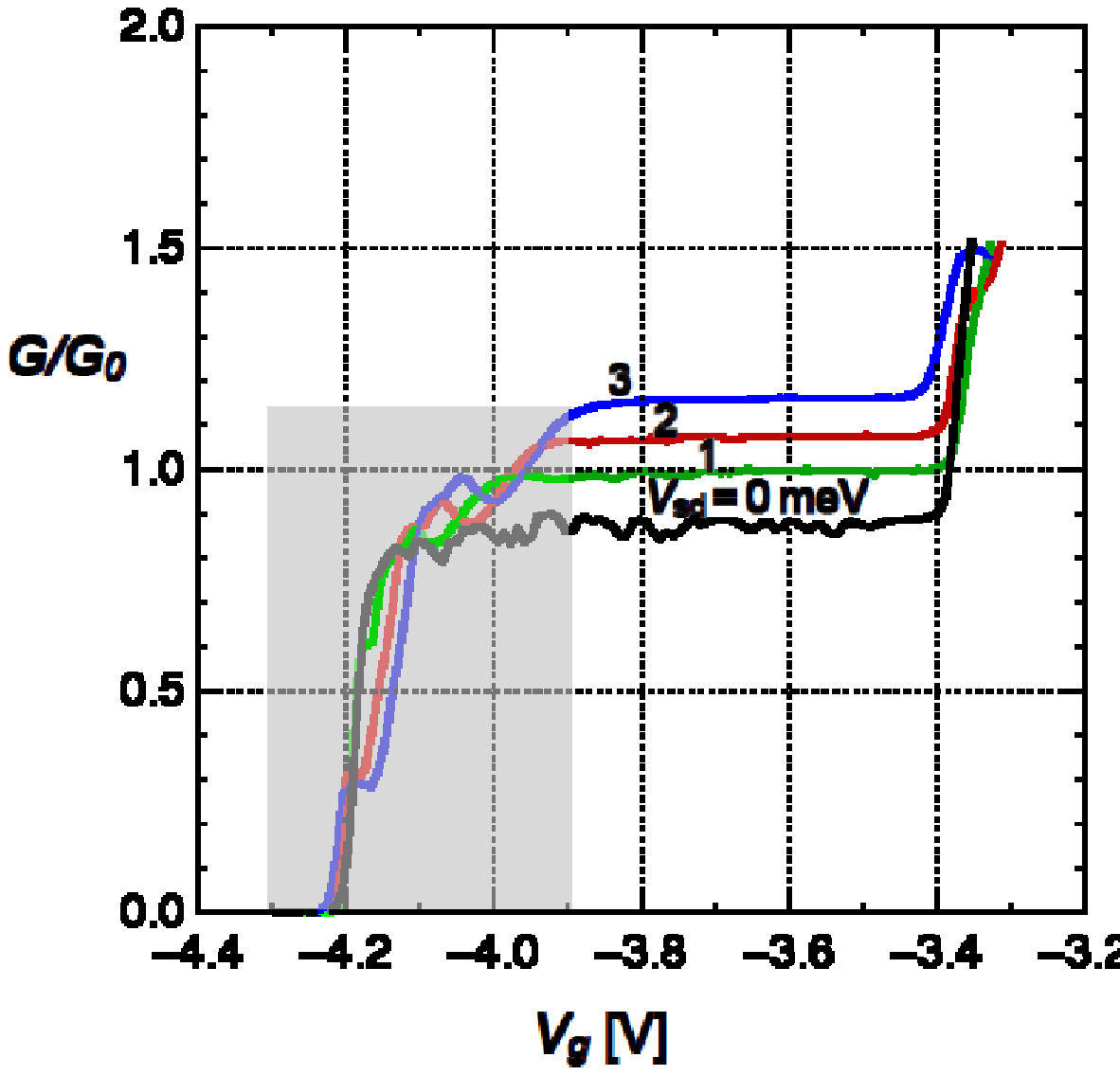}
}
{{\small {\bf FIG. 1.}
Quantized differential conductance $G$
measured in a quasi-one-dimensional multi-sub-band ballistic channel,
after de Picciotto {\em et al}
\cite{dp2}.
$G$ is plotted in units of the quantum
$G_0=77.48\mu$S, as a function of control-gate voltage $V_g$
modulating the carrier density within the channel.
Beyond the labile structures (grey box) at threshold of the
ground-state band lies a series of very flat extended plateaux
for different values of bias voltage driving the current.  
That the system is beyond linear response is seen in the
bias dependence of the conductance. The steps
of $G$ above the presumed upper limit $G_0$ is noteworthy. This
shows that effects beyond simple quantum-coherent transmission
dominate the transport physics. Adapted from
\cite{intb}.
\textcopyright ~IOP Publishing Ltd.
}}

In an investigation of the so-called 0.7 conductance
anomaly in a multi-channel QPC, de Picciotto and co-workers
\cite{1,dp2}
went further to present additional measurements of quantized
conductance in a close-to-ideal 1D device, covering the density
regime where the lowest channel's band was well occupied
right up to the threshold of the next-higher band.
When driven beyond weak bias voltages, the surprise was that
the size of the step in $G$, still essentially flat,
increased systematically and substantially with $V$.
Figure 1 shows their results.

Other than the structures greyed out in Figure 1, the absolutely
flat voltage-dependent plateaux demonstrate that more than single-particle
physics is at work. We followed up the experimenters' own hypothesis
\cite{1,dp2}
that exchanges of carriers between the lower and upper band could
enhance the conductance of the well populated lower band
by depleting it (with no effect on its conductance step size),
thereby adding excess carriers to the upper band in its threshold region
where $G$ is quite sensitive to small changes in its band density.

We reasoned that, at a higher values of $V$, the coupling between channels
leads to greater transfer of carriers to the higher level and that,
at the same time, these excited carriers have an increased likelihood
of falling back into the lower band, resulting in a steady state
regulated by negative feedback and stabilizing the enhancement of $G$
around the threshold of the upper band.

One can compare the results of our calculation in Figure 2 
\cite{intb}
with the experimental data in Figure 1.
While the calculated conductances overestimate the measured step values
in our simplified simulation, the robust character of the plateaux
and their voltage dependence are strikingly similar to experiment.
%
\vskip 0.25cm
\centerline{
\includegraphics[width=6truecm]{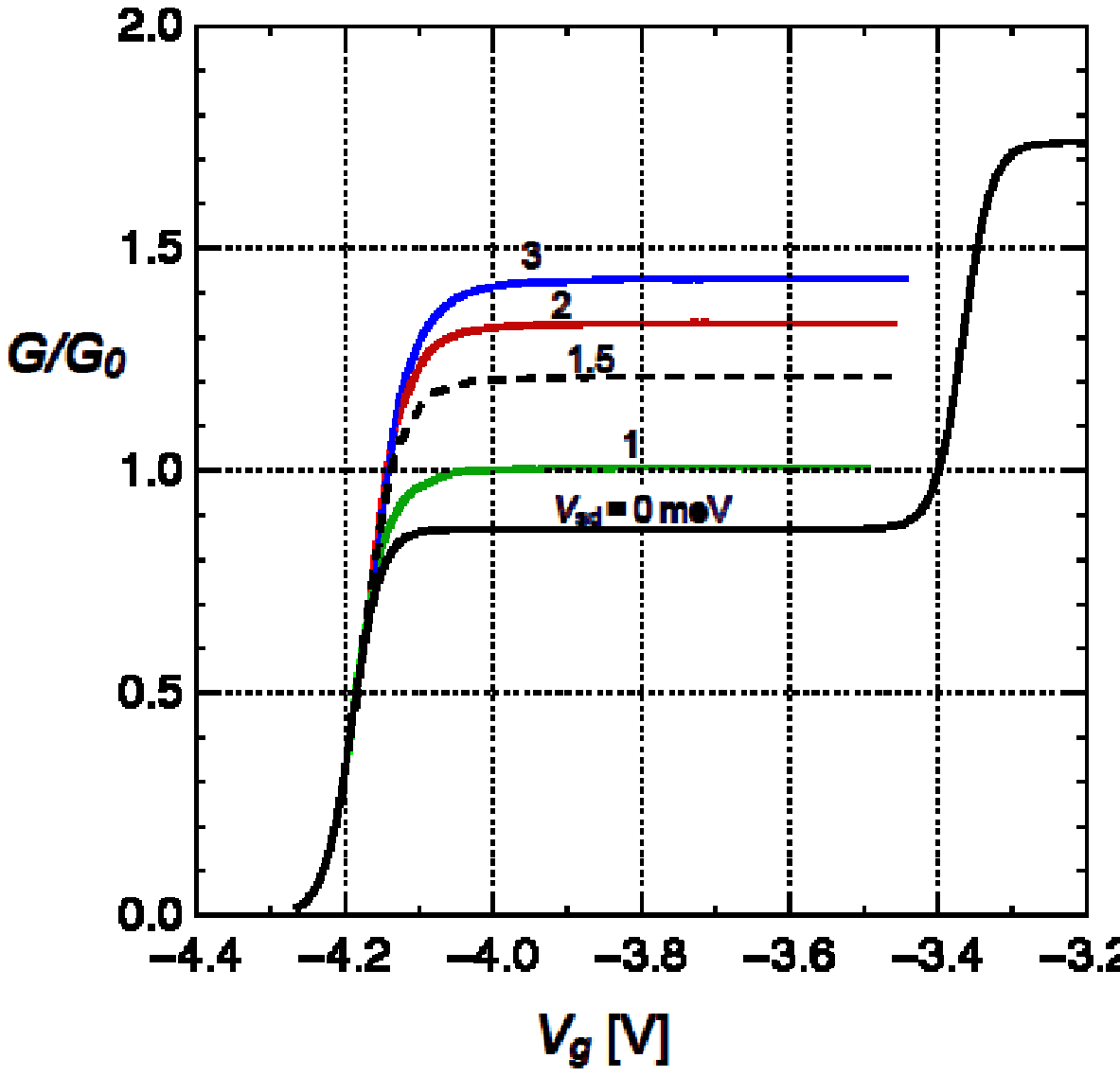}
}
{{\small {\bf FIG. 2.} {\small Anomalous enhancement of conductance
for a ballistic device model, equivalent to that of our
figure 1 (after figure 3(a) of 
\cite{dp2})
at nominal temperature 4K. Axis scales
are as for that figure; $G$ is plotted in units of $G_0$
versus gate voltage $V_g$ sweeping the electron density through the
band-gap regime from the start of the ground-state band to the
threshold of the first excited-state band.
Bottom curve: in weak-field response the quantized conductance matches
that for standard linear response. Higher curves: as the driving voltage
$V_{\rm sd}$ increases, the conductance acquires a nonlinear enhancement.
The action of interband transitions dynamically redistributes
carrier density between bands. This is responsible for the
strong enhancement of the step in $G$, beyond the upper bound
posited by quantum-transmission models of conductance.
Adapted from
\cite{intb}.
\textcopyright ~IOP Publishing Ltd.
}}
\vskip 0.25cm

The calculation involves solving a coupled quantum Boltzmann
equation including a generation-recombination mechanism
induced by the interband transition rate.
Here we set out the schematics of the procedure;
for details see 
\cite{intb}.
We restrict the problem to two channels separated by
a given band gap $E_g$. Since the total density $n(\mu)$
over both bands is constant at global chemical potential $\mu$,
conservation enforces
\begin{eqnarray}
n(\mu)
=
n_1(\mu_1) + n_2(\mu_2 - E_g)
\label{06}
\end{eqnarray}
with $n_1$ the density in the lower band and $n_2$ in the upper.
The populations out of equilibrium are no longer determined by $\mu$
but depend on the bias voltage. Therefore we refer the separate carrier
distributions to a pair of nominal equilibrium states at the effective
chemical potentials $\mu_1$ and $\mu_2$. We expect the former potential
to decrease as the lower channel depletes, and the latter to rise
as the upper channel gains extra carriers. At equilibrium both match $\mu$
but in the driven system they may not, and are nevertheless
coupled by the need to conform to conservation.

To fix $\mu_1$ and $\mu_2$ uniquely, thus closing the now
self-consistent problem, a second and physically motivated constitutive
relation is necessary.
In the parameter space of $\mu_1$ and $\mu_2$ we compute
the free energy, plotting its difference with the free energy
for totally independent bands where $\mu_1 = \mu = \mu_2$.
When there is no maximum, the solution reverts to the latter case.
If there is a nontrivial maximum in the difference,
its locus will intersect the contours of constant $n$ generated by
Equation (\ref{06}). In either eventuality one
can find the operating point at a given total density, where the net
conductance is calculated. Figure 2 is the result of our strategy.

The physical motivation for using the free energy is that it
carries the overhead of available electrical energy absorbed
from the driving field and momentarily retained as kinetic energy
of the excited carriers; a sort of reservoir in momentum space.
It is reasonable to think that this dynamic overhead builds up to
a maximum, depending on the capacity of inelastic collisions to
dissipate it and sustain a steady state. The operating point
will thus arrange itself at the free-energy maximum because the
first port of call for the transferred electrical energy is always
the carrier motion, not the heat bath which is accessed only via
the lossy collisions forming a bottleneck.

\section{QPC noise at high fields}

\centerline{
\includegraphics[height=5.5truecm]{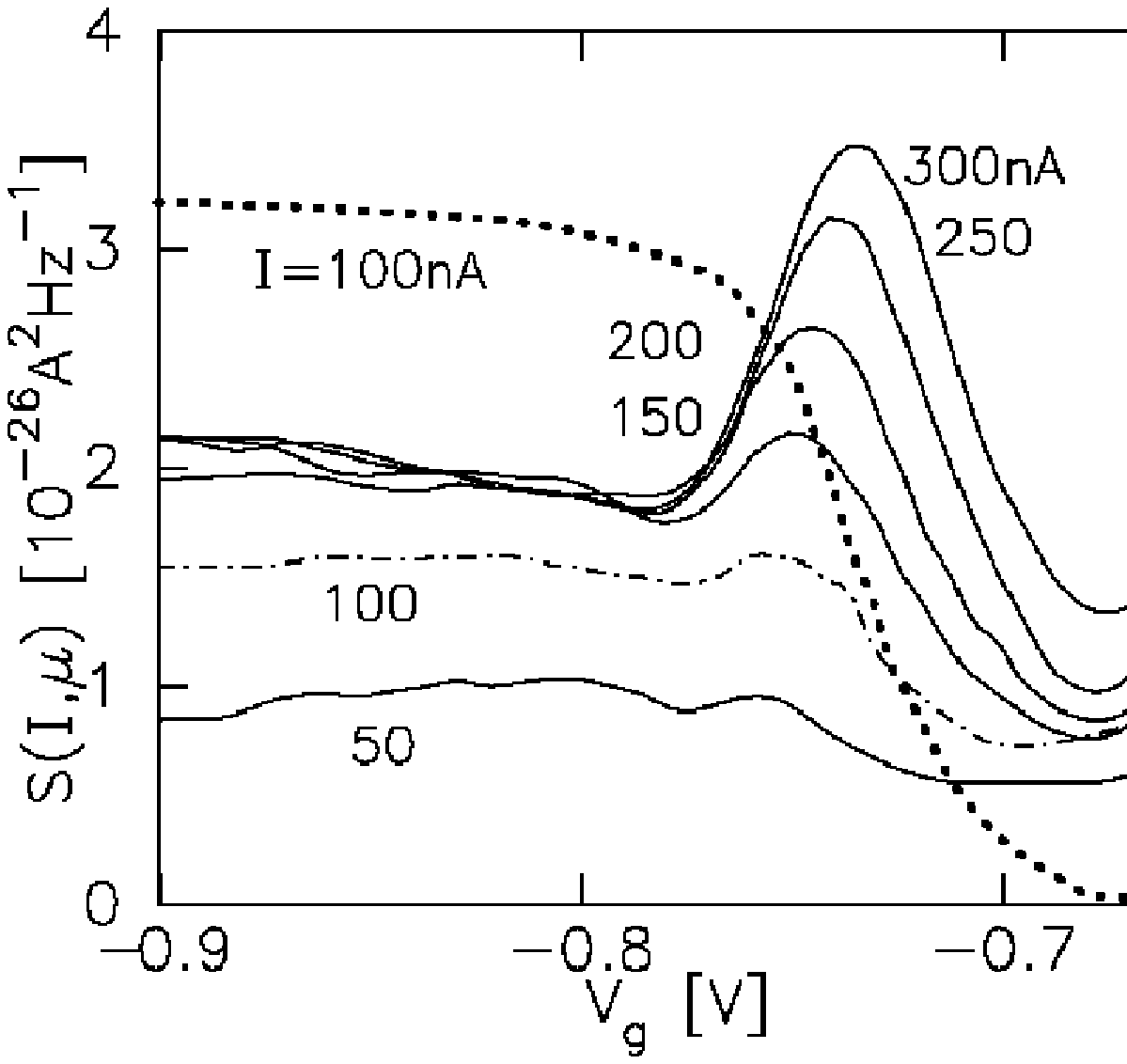}
}
{{\small {\bf FIG. 3.} Measured high-field excess current noise
in a quantum point contact at 1.5 K (adapted from 
\cite{gtd}
after Reznikov {\em et al}
\cite{2})
as a function of gate bias at fixed levels of the current.
Dotted line: the most widely adopted theoretical noise model
\cite{bb}
predicts a strictly monotonic noise signal at the first
band threshold. The very strong noise peaks actually
recorded at threshold are not anticipated.
\textcopyright ~2004 The American Physical Society.
}}
\vskip 0.25cm

Kinetic theories, such as the Boltzmann equation, are explicitly
structured to respect microscopic conservation. Subjected to
perturbation analysis, the derived equations guarantee that same
conservation properties of particle number, momentum and energy (the
latter in thermodynamic conformity with the dissipation rate)
are inherited by the corresponding fluctuations.

We have extended the kinetic approach to studying thermal current
fluctuations departing from equilibrium and manifesting as dynamical
noise in actual devices. Shot noise, thermodynamically distinct
from thermal noise, is also kinetically tractable
(other species such as $1/f$ noise require techniques that
involve explicit input about the specific environment).
For a general introduction we direct readers to Kittel 
\cite{kittel}.

Unlike fluctuations in the power output $IV$ of a conducting structure,
routinely probed in device characterization, current-current
fluctuations are not easy to measure and need specialized methods.
A thorough experimental study of high-field
current noise in QPC samples was made by Reznikov {\em et al}
\cite{2}. To highlight the anomaly in their experiment over against
commonly anticipated expectations for the density dependence of
noise, we refer to Figure 3 adapted from 
\cite{2}.

Figure 3 shows the low-frequency spectral density of excess QPC current
noise after subtraction of the normal Johnson-Nyquist thermal noise floor
$S_{\rm JN} = 4Gk_{\rm B}T$ at temperature $T$. The Reznikov group
performed their measurements at fixed levels of current through the
structure, while most other investigators choose to fix the driving
voltage only.

According to standard descriptions, at very low gate voltage (thus
carrier density) up to threshold, the signal should be dominated
by classical shot noise, determined by the Poissonian statistics of
electron injection and extraction rather than the fluctuation-dissipation
dynamics underlying $S_{\rm JN}$. As the carrier density rises within
the channel, Pauli degeneracy starts to suppress the noise, which
dies monotonically above the conduction threshold.
This is predicted for any level of the QPC current.

The anticipated uniformly monotonic fall-off of the noise is not seen,
however. Against theoretical expectation, a very strong
peak structure develops at the threshold where the lowest conduction
band of the 1D structure begins to populate. In contrast, a kinetically
grounded analysis of the fluctuations in the highly nonequilibrium
carrier distribution provides a straightforward and reasonably quantitative
account. Figure 4 displays our calculation
\cite{gtd}.

Lacking scope here for a detailed account, we offer sources
\cite{gtd}
and
\cite{spie}
as resources for a quantum Boltzmann approach to nonequilibrium
current noise. The origin of the peak structures
shown in Figure 4 lies in the interplay of the electrons'
compressibility, which falls off rapidly on
entering the degenerate density regime, and
the conductance of the QPC which decays in the depletion
region but rises at threshold to a constant fraction
of $G_0$ at large occupancy.
\vskip 0.25cm
\centerline{
\includegraphics[height=5.5truecm]{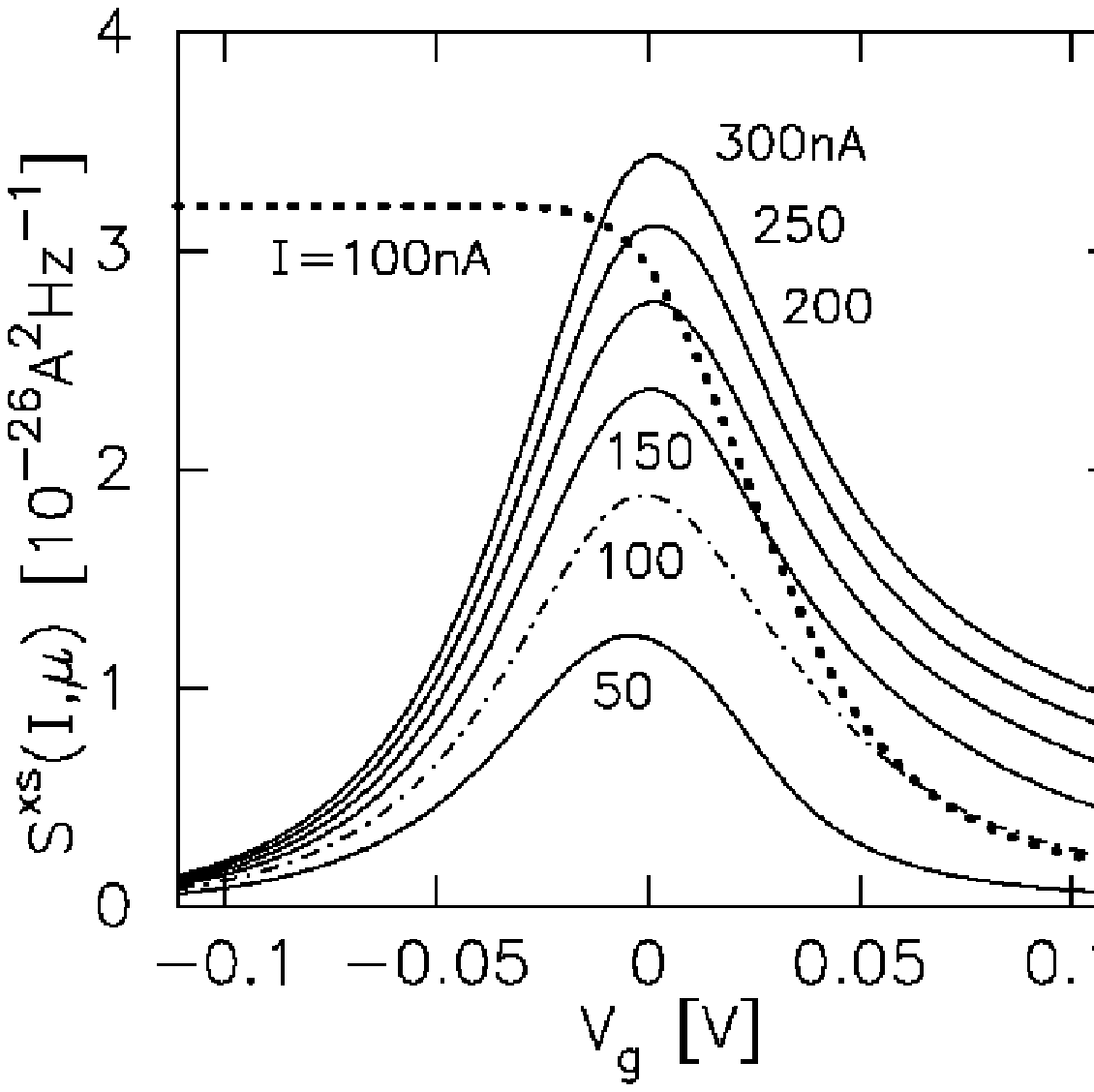}
}
{{\small {\bf Figure 4.} Computed high-field excess current noise
at 1.5 K in a QPC at fixed levels of source-drain current 
around its lowest band threshold, matching the conditions
for figure 3, after the experiment of Reznikov {\em et al}
\cite{2}.
Our calculation follows a strictly conserving quantum-Boltzmann
analysis of current fluctuations far from equilibrium
\cite{gtd}.
Note the quantitative affinity of the calculated peaks
with the experimentally observed first-threshold maxima in
figure 3. Dotted line: corresponding prediction at a
current of 100 nA, according to a quantum-transmission model
of QPC current noise
\cite{bb},
using our associated evaluation of conductance as that
model’s phenomenologically required input. This can
be compared with the peak at 100 nA (dot-dashed line)
resulting from the kinetic theoretical calculation.
\textcopyright ~2004 The American Physical Society.
}}
\vskip 0.25cm

At high current the conductance is degraded by additional
inelastic scattering. Roughly speaking we can write
the excess thermal current noise as
\begin{eqnarray*}
S^{\rm xs}(I,\mu) \sim S_{\rm JN}
\frac{\kappa(\mu)}{\kappa_0(\mu)}
{\left( \frac{G(0^+,\mu) - G(I,\mu)}{G_0} \right)}^b
\end{eqnarray*}
where the compressibility $\kappa(\mu)$ is scaled to its classical
value $\kappa_0(\mu) = n(\mu)k_{\rm B}T$
and $b$ is an exponent of order 0.5.
While $\kappa/\kappa_0$ decreases rapidly from unity in the threshold range,
the conductance, scaling with exponentially small density below that,
rises to its plateau in just that region.
The competition between the two generates the peaks at threshold.

We have accounted for the unexplained peak structures in the
experimental curves of Reznikov {\em et al}
\cite{2}
of Figure 3. So far we have not advanced a similar description
of the shot-noise-like plateaux seen there at gate voltages below -0.8V,
corresponding to extremely low carrier numbers and high driving fields.
In the predicted versus measured 100nA curves in
that Figure, standard models
\cite{bb}
overestimate measured levels but contain no further
mechanism to correct the predicted excess.
The issue is open to a future explicit account
of the shot-noise aspect, though from a kinetic standpoint.

Finally, it is noteworthy that the compressibility central to
our noise theory does not depend in any way on the size of the current
through the QPC; its value is determined
by the global chemical potential alone. This is due to
strong metallic screening (and rapid thermal dissipation)
from the large, degenerate population of free electrons in
the macroscopic device leads with the overall device neutrality that
follows from it.
\cite{2000}.

In physical terms the total number $N$ of carriers in the working region
(metallic channel plus interfaces), and its mean thermal fluctuation
$\Delta N = k_{\rm B}T\delta N/\delta \mu$, are environmentally fixed;
the electrostatic boundary conditions constrain $N$ and $\Delta N$
to their equilibrium values set globally by $\mu$ irrespective
of driving voltage.
Any internal redistribution of the driven carriers cannot manifest
in the bulk compressibility of the active structure
\cite{2000}.
This is the direct outcome of the physics of charged Fermi liquids
\cite{pines};
a fundamental collective effect insufficiently considered by
treatments of 1D fluctuations and violated by some
\cite{ijmpb},
even to a loss of gauge invariance (charge conservation)
\cite{bb}.

\section{Nonlinear Aharonov-Bohm effect: a challenge}

Our last example presents a fascinating theoretical challenge whose
resolution in terms of standard quantum kinetics remains wide open.
The Aharonov-Bohm effect has long been understood as a demonstration
of the physical reality of the vector potential
\cite{baym}.
The way by which the vector potential enters directly into the
wave function of a carrier results in a relative phase shift
when the carrier has the option of passing to one or other side of a
tubular region, enclosing a magnetic flux.

Even if the magnetic field were screened to vanish outside its confining tube,
the generating vector potential would not be zero, acting differently on the
phase of each branch of the carrier's wave. This provides a variation on
the two-slit interference scenario: modulating the magnetic flux induces
a phase mismatch between branches, causing a periodic pattern of
oscillations in the magnitude of the current.

Two significant experimental investigations of nonlinearity in the
Aharonov-Bohm effect were conducted by Neder {\em et al}
\cite{3}
and by Leturcq {\em et al}
\cite{3.1}.
For brevity here we linit ourselves to discussing the former results.
In the configuration of Reference
\cite{3}
the difference of maximum (constructive interference)
and minimum (destructive interference) current magnitudes,
as a ratio with their sum,
defines the visibility: this measure lies between one and zero
as the magnetic flux is varied. The investigated structure
consisted of two QPCs linked in parallel to make up a Mach-Zehnder
electron interferometer enclosing an area threaded by a magnetic field.
Its flux was modulated by electrostatically
modifying one of the QPC branches, altering both its path length
and the enclosed area.

For low driving voltage the visibility should be constant at fixed magnetic
field, since all currents scale linearly with voltage. Away from the linear
region, based on previous experience with a simpler device, the
investigators anticipated a progressive monotonic decay in visibility
owing to loss of quantum coherence through increased random scattering
\cite{3}.

The Neder collaboration found, instead, that the visibility as a function
of greater driving voltage first fell rapidly to zero but then recovered
and rose substantially, repeating this behavior with a periodicity
strictly correlated with an abrupt change by $\pi$ in the
stepwise form of the overall phase difference across the interferometer,
inferred from the data. As carefully discussed in their paper
\cite{3}
the effect contradicts conventional explanations.

At this writing we have no light to shed on this remarkable finding.
We are, however, intrigued by the robust periodicity of the visibility  
and consider that it well merits further thought, given that
nonequilibrium transport in most cases really should wipe out such
delicate effects. One potential consideration may be relevant: a model
of the twin-path electron interferometer should perhaps not assume ideal,
cost-free 50-50 transmission at its junctions as for, say, half-silvered
mirrors in its optical predecessor. Rather, the hole in the conductive
geometry presents a different sort of quantum barrier to an incoming
carrier, not just dividing but severely distorting its wave function.
The barrier is not fully 1D but ``one-plus-one-dimensional'',
calling for detailed simulation to compute how the wave function
negotiates being split, to heal itself after passage.

Calculations show that normal quantum transport
through multiply connected structures results in transmission factors
with a complex energy dependence, more so as these geometries
cause states in discrete bands to hybridize, complicating the eigenstate
solutions. Such transmission behaviour, inherent in this topology,
would likewise strongly condition the carrier flux. How this might
play out for any of the finer details of the Aharonov-Bohm phase shift
\cite{3},
with carriers in different bands possibly coupled self-consistently and
reacting differently to the barrier, is certainly not clear.

\section{Summary}

We have offered a brief recapitulation of several past measurements of
anomalous one-dimensional conductance, and of its closely associated
current noise. In their quite different ways they have all provided a
strong incentive to look for theoretical understandings of nonequilibrium
1D transport. In order of appearance we have covered: enhancement
of nonlinear conductance beyond the Landauer bound, in the energy gap
of a multi-channel device; the unexpected noise peak in a strongly
driven quantum point contact; and, still not satisfactorily explained,
the remarkable periodic quenching and resurgence of Aharonov-Bohm current
oscillations out of equilibrium.

From all these cases the message is that, beyond phenomenology, a
microscopically based many-body methodology is needed for nonequilibrium 1D
transport; a toolbox equipped to move, in a controlled way, past
single-particle descriptions essentially predicated on tunnelling
at linear response.
Such approaches have always been possible thanks to the massive extant
literature on quantum kinetic theory.
\cite{mahan,pines,pcm,frensley2,rammer,roepke,bonitz}
In our own studies we have made
honest attempts to adapt this solidly founded body of work to the
one-dimensional realm, frequently in simplified models but always respecting
conservation.

This brings us to a message gleaned from our own efforts. It is indeed the
microscopic conservation laws, including gauge invariance, that shape
the solution to problems of 1D charge transport both near and far from
equilibrium, often to a surprising extent (a notable example is the
previously unsuspected role of bulk electronic compressibility in
high-field thermal current noise). While not sufficient on their own to
solve these problems, they are absolutely indispensable.


%
\end{document}